\documentclass[aps,prl,superscriptaddress,twocolumn,preprintnumbers,
nofootinbib,showpacs]{revtex4-2}
\pdfoutput=1

\usepackage{graphicx}
\usepackage{epsfig}
\usepackage{amssymb}

\usepackage{amsmath}
\usepackage{amsfonts}
\usepackage{fancyhdr}
 
\usepackage{amsmath}
\usepackage{anysize}
\usepackage{verbatim}
\usepackage{float}
\usepackage{cancel} 
\usepackage{framed}
\usepackage[titletoc]{appendix}
\usepackage{mathrsfs}
\usepackage{amsbsy}

\usepackage{vmargin}

\usepackage{bm}

\usepackage{color}

\setpapersize{A4}
\setmargins{2.5cm}       
{1.5cm}                        
{17cm}                      
{23.42cm}                    
{10pt}                           
{1cm}                           
{0pt}                             
{2cm}                           

\newcommand{\be}{\begin{equation}}
\newcommand{\ee}{\end{equation}}
\newcommand{\bea}{\begin{eqnarray}}
\newcommand{\eea}{\end{eqnarray}}
\newcommand{\nn}{\nonumber}

\begin{document}

\title{High-Temperature QCD Static Potential beyond Leading Order}

\author{Margaret E. Carrington}
\affiliation{Department of Physics, Brandon University,
Brandon, Manitoba R7A 6A9, Canada}
\affiliation{Winnipeg Institute for Theoretical Physics, Winnipeg, Manitoba, Canada}
\author{Cristina Manuel}
\affiliation{Instituto de Ciencias del Espacio (ICE, CSIC) \\
C. Can Magrans s.n., 08193 Cerdanyola del Vall\`es, Catalonia, Spain}
\affiliation{Institut d'Estudis Espacials de Catalunya (IEEC) \\
08860 Castelldefels (Barcelona), Catalonia, Spain
}
\author{Joan Soto} 
\affiliation{Departament de F\'\i sica Qu\`antica i Astrof\'\i sica and Institut de Ci\`encies del Cosmos, 
Universitat de Barcelona, Mart\'\i $\;$ i Franqu\`es 1, 08028 Barcelona, Catalonia, Spain}
\affiliation{Institut d'Estudis Espacials de Catalunya (IEEC) \\
08860 Castelldefels (Barcelona), Catalonia, Spain
}

\date{\today}

\preprint{}

\begin{abstract}

We calculate the leading and next-to-leading corrections to the real-time QCD static  potential in a high temperature medium  in the  region where bound states transit from narrow resonances to wide ones. We find sizable contributions to both the real and the imaginary part of the potential. The calculation involves both loop diagrams calculated in the Hard Thermal Loop (HTL) effective theory and power corrections to the HTL Lagrangian calculated in QCD. We compare our results with recent lattice data and check the consistency of different methods used in lattice calculations. We also discuss the usefulness of our results to guide lattice inputs.

\end{abstract}

\maketitle

{\bf 1. Introduction.}
In quantum chromodynamics (QCD) the static potential provides the dominant interaction between a heavy quark and a heavy antiquark at low energy.
At zero temperature it is attractive, Coulomb-like at short distances and linearly rising at long distances, a sign of confinement \cite{Wilson:1974sk}. At high enough temperature the QCD coupling constant is expected to be small and perturbative calculations of the static potential are realistic. The short-distance Coulomb-like potential becomes screened, as in quantum electrodynamics, namely Yukawa-like, with a screening mass proportional to the temperature. As the temperature increases the Yukawa-like potential supports fewer and fewer bound states until at high enough temperatures there would no longer be any bound-states. The suppression of heavy quark-antiquark bound state production in Heavy Ion Collision (HIC) experiments through screening
appeared to be a solid prediction of QCD and a reliable signal for quark-gluon plasma (QGP) formation \cite{Matsui:1986dk}.
However a closer inspection of the static potential not only revealed that it develops an imaginary part \cite{Laine:2006ns}, but also that the imaginary part is parametrically larger than the real part when the screening effects become important \cite{Escobedo:2008sy}. This implies that the disappearance of bound states occurs because they become wide resonances rather than because they are no longer  supported by the Yukawa-like potential. The suppression of heavy quark bound state production  is indeed observed in current HIC experiments, clearly in the bottomonium ($\Upsilon$) family \cite{ALICE:2020wwx,CMS:2018zza} and indirectly for the charmonium one \cite{CMS:2017uuv}, as recombination effects must be taken into account for the latter. Experimental results show that the peaks corresponding to bottomonium states become wider and fuse with the background as the multiplicity (temperature) increases, rather than sequentially disappearing with roughly the same width. This pattern is consistent with the idea that bound states disappear because they decay, and not because the potential is screened to the point that it becomes too shallow to support them.

For most of the temperatures  reached in HIC experiments the QCD coupling constant may not be small enough to trust perturbative estimates. An enormous effort has been made in the past decade to identify both the real and the imaginary part of the real-time static potential from non-perturbative lattice QCD calculations \cite{Rothkopf:2011db,Burnier:2014ssa, Burnier:2015tda, Burnier:2016mxc, Lehmann:2020fjt, Boguslavski:2021zga, Bala:2021fkm,Dong:2022mbo,Bazavov:2023dci}. This is a difficult task, both conceptually and technically. Early attempts to identify the static potential from several objects related to Polyakov loop correlators missed the imaginary part \cite{Kaczmarek:1999mm,Kaczmarek:2002mc,Petreczky:2004pz,Kaczmarek:2005ui,Maezawa:2007fc, Burnier:2009bk}. The extraction of the real-time static potential from Euclidean correlators calculated on the lattice 
requires Bayesian methods, which need a physically motivated guess for the form of the spectral function \cite{Jarrell:1996rrw,Asakawa:2000tr, Rothkopf:2011ef, Burnier:2013fca, Burnier:2013nla}. The perturbative result is often used as a guess. 

We calculate the real-time QCD static potential  beyond leading order (BLO) in a high temperature medium where bound states start melting, namely where they transit from narrow to wide resonances. The purpose of this calculation is twofold. We check the robustness of the leading order result on which our qualitative understanding is based. We also put forward a wider set of physically motivated forms of the potential which can be used as an input for the Bayesian methods to extract the real-time static potential from Euclidean correlators calculated on the lattice.

The static potential is defined as the ground-state energy of a static quark and a static antiquark separated at a distance $r$. If the quark and antiquark are linked by a straight Wilson line, to get a gauge invariant physical state, the calculation amounts to that of the expectation value of a rectangular Wilson loop with the length of the temporal sides of the loop  taken to infinity. This leads to the coordinate space potential. Alternatively one can use standard diagrammatic methods \cite{Brambilla:2004jw,Pineda:2011dg} which give the momentum space potential. We have carried out the calculation both ways. 
We work in the close-time-path formalism of thermal field theory and we use an approach based on the Hard-Thermal-Loop (HTL) effective theory \cite{Braaten:1991gm}.   The static quark and antiquark are (unthermalised) probe particles for which only the longitudinal photon ($A_0$) vertices in the time-ordered branch are relevant. 
The vertices in the anti-time-ordered branch will only enter in the calculation of the gluon self-energy (discussed below). We use the Coulomb gauge and dimensional regularization throughout, $C_F=(N_c^2-1)/(2N_c)$ where $N_c$ is the number of colours, and $g$ is the QCD coupling constant. 

The leading order (LO) result for the momentum space potential under the assumptions that $g\ll 1$ and that the typical momentum exchange between the quark and antiquark satisfies $p \ll T$ is \cite{Laine:2006ns}
\bea
&& V_{1{\rm LO}}(p)= g^2 C_F G(0,p)   \ , 
\label{lo-1} \\
&& G
(0,p)=-\frac{1}{m_D^2+p^2}+ \frac{i \pi  T m_D^2}{p \left(m_D^2+p^2\right){}^2}\nonumber
\eea
where $m_D = gT \hat m_D$ is the Debye mass, with $\hat m_D = \sqrt{(N_c+N_f/2)/3}$, $T$ is the temperature, $N_f$ is the number of light flavors, and $G(p_0,p)$ the time-ordered longitudinal HTL propagator. 
The corresponding coordinate space potential is 
\bea
V_{1{\rm LO}}(r) = g^2 C_F \int \frac{d^3p}{(2\pi)^3}\, e^{i \vec p\cdot\vec r } \, G(0,\vec p)\, ,
\label{v1-int}
\eea
where $r=\vert \vec{r} \vert$.
Performing the Fourier transform one obtains
\be
 V_{1{\rm LO}} (r) =   -  \frac{ g^2 C_F  }{4\pi {\hat r}} \left(m_D e^{-\hat r } - 2 i T \, I_2(\hat r) \right)
\ , 
\label{mikko-real-2}
\ee
where we define the dimensionless integrals
$I_j(\hat r) = \int_0^\infty d\hat p \, \sin \left(\hat{p} \hat{r}\right)(\hat{p}^2+1)^{-j}$, and
$\hat r = r m_D$. 

At any order in the loop expansion of the expression that defines the potential, we must decide if and how to dress the propagators and vertices in the resulting momentum integrals (i.e. if we should use bare $n$-point functions, or HTL corrected ones). This decision depends on the momentum scale that we choose to focus on, which in turn will determine the range of $r$ where our coordinate space potential will be most reliable. 
To decide what momentum scale is important for our purposes we return to Eq.~(\ref{lo-1}). 
For a narrow resonance to exist the imaginary part of the static potential must be smaller than the real part. This implies that $p>(m_D^2T)^{1/3}\sim g^{2/3} T$ or $p=g^a T$ with $0<a<2/3$. 
Note that since $T \gg p \gg m_D$ we can simultaneously satisfy the conditions that perturbation theory is valid and that narrow resonances exist. This hierarchy of scales could be exploited using non-relativistic effective field theories \cite{Caswell:1985ui,Pineda:1997bj,Brambilla:2008cx}, but we  use instead the technique of integration by regions \cite{Beneke:1997zp,Smirnov:2012gma}, which is better suited for our purposes. We will describe the scale of the momentum $p$ as semi-hard. 
The value $a=2/3$ gives the momentum scale for which the real and imaginary parts of the leading order  momentum space potential are of the same size and is parametrically the scale at which we expect quarkonium to dissociate. We will take into account all corrections larger than $g^2$ to the real part and larger than $g^{3a}$ and $g^{2-a}$ to the imaginary part of the LO momentum space potential. Note that the scaling in $a$ at LO is different for the real ($g^{2-2a}/T^2$) and the imaginary parts ($g^{4-5a}/T^2$). Two loop diagrams give corrections of order $g^2$ or smaller to the real part, and of order $g^{3a}$ or smaller to the imaginary part. Including full HTL vertices in the self-energy diagrams gives corrections of order $g^2$ or smaller to the real part and of order $g^{2-a}$ or smaller to the imaginary part. All together our conditions restrict $a$ to be in the range $1/3\,<\,a\,<\,2/3$. 
The coordinate space potential that we obtain upon Fourier transforming the momentum space potential will not be valid for $r m_D\gg 1$, since large $r$ corresponds to momenta softer than the semi-hard scale we have chosen, or  $r T\ll 1$, 
since small $r$ corresponds to momenta larger than the temperature. 
We will extend the region where our coordinate space potential is valid by Fourier transforming a momentum space expression with all denominator factors of the form $p^2+m_D^2$ unexpanded. 
Since we have assumed $p\gg m_D$ this procedure is not fully consistent but it 
extends to larger $r$ the region where we expect our result to be valid. 
The contributions to the coordinate space potential from the soft region ($p\sim m_D$) for $m_D\ll 1/r\ll T$ that we do not calculate have a universal form at any order in $g$. Since we assume $rm_D\ll 1$, the exponential $e^{i\vec p\vec r}$ in the Fourier transform can be expanded in the soft region. The contribution from this region thus reduces to a polynomial in $r^2$ (which may depend on the factorization scale
)\footnote{For anisotropic plasmas, it becomes a polynomial in $\vec r$. }.
At the order to which we work we need the zeroth order (the $r$-independent term) for the real part and up to the first order (the $r^2$ term) in the imaginary part \cite{Supplemental}. We note that the contributions from the heavy quark self-energies in the static limit add to the zeroth order terms just mentioned. 

\begin{figure}[H]
\begin{center}
\includegraphics[width=0.45\textwidth]{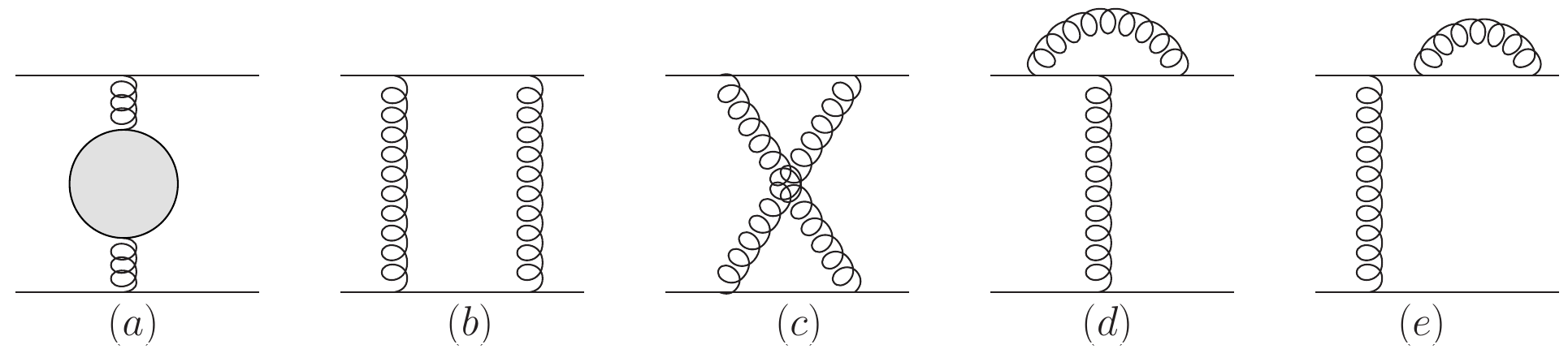}
\caption{One-loop contributions to the static potential in the Coulomb gauge. All gluon lines correspond to longitudinal gluons. The iteration of the LO potential must be subtracted. Note that there is no diagram analogous to (d) with a three-gluon vertex.
\label{fig-diag}}
\end{center}
\end{figure}

{\bf 2. The static potential beyond leading order. }
The relevant diagrams are shown in Fig.~\ref{fig-diag}, where all the fermion lines do not depend on the spatial momenta  (the static limit). There is one (4-dimensional) momentum variable that is integrated over (which we call $k$) and the momentum transfer is $p=(p_0,\vec p)$. The energy transfer $p_0$ is eventually taken to zero to get the momentum space potential. Fourier transforming the momentum space potential gives the coordinate space potential. 
There are two relevant contributions to the gluon self-energy $\Pi^{\rm ret}_{\rm BLO}$ (first diagram in Fig.~\ref{fig-diag}). One comes from the power correction to the HTL self-energy ($k\sim T$), and one from the one loop self-energy diagram with the loop momentum  semi-hard $k\sim p$. The fact that the semi-hard scale dominates the soft one ($m_D$) produces enormous simplifications. Both HTL vertices and HTL propagators reduce to the corresponding bare ones up to corrections of order $m_D^2/p^2$. 
There are no contributions to $\Pi^{\rm ret}_{\rm BLO}$ from quark loops for $k\sim p$ because at energies smaller than $T$ they are Pauli-blocked and hence suppressed by $p^2/T^2$ with respect to the Bose-enhanced gluonic contributions. Also, possible contributions from tadpole diagrams vanish. 
The last four graphs in Fig.~\ref{fig-diag} are zero at zero temperature in Coulomb gauge because they involve only the bare longitudinal gluon propagator, which is energy independent. 
At finite temperature they give a non-vanishing contribution that has not previously been considered. 
The leading contribution from the last four graphs in Fig.~\ref{fig-diag} arises when the internal momentum $k\sim m_D$ and HTL propagators are used, because the HTL longitudinal gluon propagator is energy dependent. 

The gluon self-energy bubble in Fig.~\ref{fig-diag}(a) at next-to-leading order in $p/T$ is\footnote{The result in Feynman gauge is different but we have checked  that in this gauge there are additional contributions from the vertex diagrams that combine to give the same total 
contribution to the potential. }, for $p_0 \rightarrow 0$,
\bea
&& \Pi^{\rm ret}_{\rm BLO}(p_0,\vec p) \label{selfenergy} \\
&& = -\frac{g^2T}{4}\left(N_c\left(p+ i \, \frac{7}{3} \, \frac{p_0}{\pi}\right)  - i \left(N_c-\frac{N_f}{2}\right) \frac{p_0\,p}{2\pi T} \right)\,.\nonumber
\eea
The first contribution comes from the loop with $k$ semi-hard \cite{Shi:2015tmz,Zhu:2015edf}\footnote{The real part was obtained earlier in \cite{Rebhan:1993az}.} and the second piece is the power correction.\footnote{There is also a contribution from the power correction to the real part $\sim g^2 p^2$ which is not kept because it is higher order. The power correction was calculated for QED in \cite{Manuel:2016wqs,Carignano:2017ovz} and for QCD in Feynman gauge in \cite{Ekstedt:2023anj,Gorda:2023zwy}.} 
To find the corresponding contribution to the potential the result in (\ref{selfenergy}) is used in the time ordered propagator. 
Note that the terms in Eq.~(\ref{selfenergy}) proportional to the energy transfer $p_0$ must be kept, even though we take the limit $p_0\to 0$ in the calculation of the static potential, because they combine with a Bose-enhancement factor $\sim T/p_0$ in the  time-ordered propagator. 
The sizes of the contribution (\ref{selfenergy}) to the real [imaginary] part of the static potential are  $g^{4-3a}/T^2$ [$g^{4-4a}/T^2$, $g^{4-3a}/T^2$] where the two terms in the square bracket indicate, respectively, the semi-hard loop and the power correction. 

In the diagrams in Figs.~\ref{fig-diag}(d),(e) the loops are not sensitive to the momentum transfer $p$. 
The contribution 
in momentum space is
\cite{Supplemental}
\bea
V_{2}^{\rm (de)}
= ig^4 N_c C_F G(0,p) \int \frac{d^4k}{(2\pi)^4}\frac{G(k_0,k)}{(k_0+i\eta)^2}\,,
\eea
with $\eta\to 0^+$. To do the momentum integral we use the time ordered propagators in the HTL limit 
and expand the Bose distribution since $k\sim m_D$. This gives 
\bea
V_{2}^{\rm (de)}
= \frac{g^4 N_c C_F}{8\pi} G(0,p)\left(\left(1-\frac{3\pi^2}{16}\right)\frac{T}{m_D}
\right)\,.
\label{vert-wave-2}
\eea
The contribution from the graphs in Figs.~\ref{fig-diag}(b),(c) is 
\bea
\!\!\!\!\!V_{2}^{\rm (bc)}
= -\frac{ig^4 N_c C_F}{2} \!\int\! \frac{d^4k}{(2\pi)^4}\frac{G(k_0,\vec k + \vec p) G(k_0,k)}{(k_0+i\eta)^2} \,.  
\label{ladd-cross}
\eea
Using $p\gg k, k_0$ we can write, under the integral sign, 
\bea
&& G(k_0,\vec k + \vec p) =
G(0,p) \Big[1-\frac{k^2}{3}G(0,p) \left(1+ \right. 
\nonumber 
\\
&&+ \left. 
4m_D^2G(0, p) \right) \Big ]  +\mathcal{O}(\frac{m_D^3}{p^5})\,.
\label{damped}
\eea

We keep $m_D^2$ in the denominators even though it is parametrically smaller than $p^2$ and we call this the damped approximation. The dominant term in (\ref{damped}) gives $V_{2}^{\rm (bc)} (p) = -V_{2}^{\rm (de)} (p)/2$. The subdominant term only contributes to the real part of the potential.
From Eq.~(\ref{vert-wave-2}) we see that the real part of 
Fig.~\ref{fig-diag}(b),(c)  is $\sim g^4 T/(m_D p^2)$ and is parametrically larger than the contribution from the self-energy diagrams (\ref{selfenergy}). We note that when the loop momentum is $k\sim p$ it appears that we could get a contribution from Fig.~\ref{fig-diag}(b),(c)  of the same size as the self-energy contribution, but  in this case the HTL propagators can be approximated by the corresponding bare ones and therefore we get zero in Coulomb gauge. 
For reference we give the result obtained from
Eqs.~(\ref{selfenergy}),(\ref{vert-wave-2}),(\ref{ladd-cross}), by expanding in $m_D/p$ and dropping terms of order $g^4T m_D^2/p^5$ in the real part and ($g^4 T^2m_D^2/p^6$, $g^4/p^2$) in the imaginary part. This result 
reads
\bea\label{expanded}
V_{2, {\rm exp}}(p)&=&-\frac{g^4C_F N_c T}{16\pi m_D p^2}\left\{
\left[1-\frac{3 \pi ^2}{16} +  \frac{4 \pi m_D}{p}\right.\right. \\
&& \left.\left.+ \frac{m_D^2}{p^2}\left(\frac{5\pi ^2}{24}-\frac{4
   }{3}\right)\right] 
	+i\frac{\pi T m_D}{ p^2} \left[ 
\frac{56}{3 \pi} \right.\right. \nonumber\\
&&\left.\left.-\left(1-\frac{3 \pi ^2}{16}\right) \frac{m_D}{p} 
-\left(1-\frac{N_f}{2N_c}\right)\frac{4 p}{\pi T}\right]\right\}\,.\nonumber
\eea
The coordinate space potential $V_{2, {\rm exp}}(r)$ obtained from (\ref{expanded}) 
has IR poles in DR \cite{Supplemental}, 
which should cancel when the soft contribution is added.
The damped approximation reproduces (\ref{expanded}) when expanded in $m_D^2/p^2$. The coordinate space potential in the damped approximation is IR finite and reads,
\bea
&& {\rm Re}[V_{2}]  =   \frac{g^4 N_c C_F   T }{64 \pi ^2 {\hat r} }  \bigg\{  8 \left(I_2(\hat r) -  I_1(\hat r) \right)
\nonumber \\
 &+ &
\frac{e^{-\hat{r}}}{16} \bigg( 3 \pi^2-16+\frac{\hat r}{6} \left(16-\pi ^2\right) \bigg) \bigg\} , 
\label{extra-1}
\eea
\bea
&& i {\rm Im}[V_{2}] = - i\frac{g^3 C_F \,T\,}{16\pi^2  \hat{m}_D} \bigg\{\frac{3 \pi ^2-16}{32\, {\hat r}}  I_2( \hat{r}) \label{extra-2} 
 \\
&&+ \frac{7}{3} \; N_c e^{-\hat r}
- \frac{2g\hat{m}_D}{\pi \hat r} \left(N_c-\frac{N_f}{2}\right) \big(I_1(\hat r)-I_2(\hat r)\big) \bigg\} \ .
\nonumber
\eea
 The first (second) line of the real (imaginary) part comes from Fig.~\ref{fig-diag}(a) and the remaining contributions from the last four diagrams in Fig. \ref{fig-diag}.

As explained previously,  we have not calculated the soft contributions to the momentum space potential. 
At the order we are working, the general form of these contributions in the coordinate space potential is 
\be \label{soft}
V_{2,{\rm soft}}=g^4 q_0 T
+ i\left(g^3 i_0 T+ g^5 i_2 r^2 T^3\right)\,,
\ee 
where $(q_0,i_0,i_2)$ are real constants, which may depend on the factorization scale. Note that these constants have the correct form to absorb the IR poles in the Fourier transform of (9) \cite{Supplemental}.

\begin{figure}[h]
\begin{centering}
\includegraphics[scale=0.68]{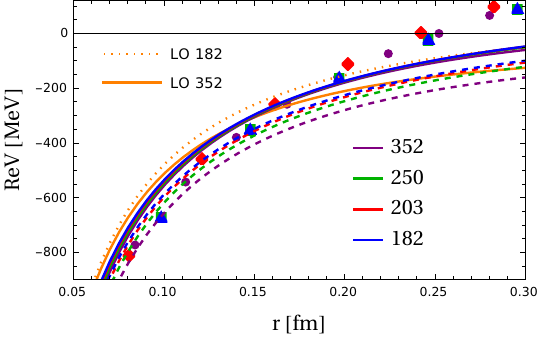}
\includegraphics[scale=0.68]{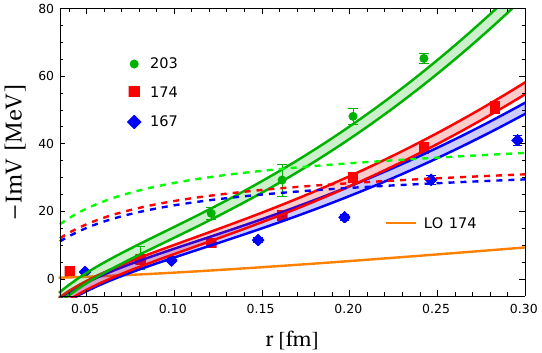}
\caption{Real and imaginary part of $V$. Dashed (solid) lines correspond to $V=V_{1{\rm LO}}+ V_2$ ($V= V_{1{\rm LO}}+ V_2+V_{2,{\rm soft}}$) in the damped approximation with parameters given in the text. The legends indicate the temperature in MeV. The real part includes a global shift of $S_{\rm avg}=209.5$ MeV for all curves. The LO contribution (HTL) includes the one-loop static quark self-energies.  For the imaginary part, 
we show a single temperature since the remaining two would overlap with it.
 The solid bands on the lower plot indicate the uncertainty in the values of the fitted constants inherited from the error bars of the lattice data. \label{real-fit}}
\end{centering}
\end{figure}

{\bf 3. Comparison with lattice results.}
We consider bottomonium, and take the mass of the bottom quark $m_b=4676$ MeV.
We compare our results with two different lattice calculations. 
First we compare our potential with \cite{Bazavov:2023dci}. We fix $g$ from the fit to the $T=0$ lattice data for the static potential with $r\in [0,0.3]$ fm which delivers $g=1.8$\footnote{The running of $g=g(T)$ is due to the power corrections to the real part of the potential \cite{Manuel:2016wqs,Carignano:2017ovz}, which are higher order in our counting ($g^2$ suppressed).}
, and we use $N_c=N_f=3$. The constants 
$(q_0,i_0,i_2)$ are determined by fitting to the lattice data at all available temperatures and values of $r$ between 0.04 and 0.3~fm.  In order to adjust the origin of energies, we also add a temperature independent constant 
$S$ to the real part of the potential. The values we obtain are 
$(q_0,i_0,i_2)=(0.027,-0.019\pm 0.001, 0.194\pm 0.002)$ and $S=219\, {\rm MeV}$. We also adjust the origin of energies for the LO potential and obtain $S_{\rm LO}=200$ MeV. We show our results in Fig. \ref{real-fit}  for $S_{\rm avg}=(S+S_{\rm LO})/2=209.5$ MeV. If we ignore the soft contributions ($q_0=i_0=i_2=0$), our results are still closer to data than the LO result, although the improvement is marginal. 
When we include them, we observe that the real part of the potential varies very little with temperature, in agreement with the data. This is non-trivial since the soft contribution is independent of $r$. The slightly different shape with respect to the data is expected to be fixed by higher order temperature independent radiative corrections, see Ref.~\cite{Bazavov:2012ka}.\footnote{A soft contribution $\sim g^6 r^2 T^3$ improves the description of data but we do not include it because it is higher order in our counting.} For the imaginary part of the potential most of the contribution comes from the soft region, which is responsible for the large correction with respect to the LO result.

\begin{figure}[h]
\begin{centering}
\includegraphics[scale=0.68]{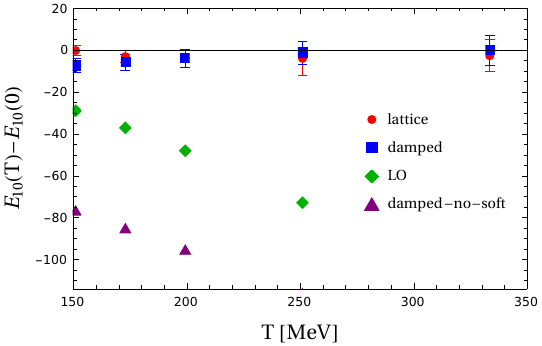}
\includegraphics[scale=0.68]{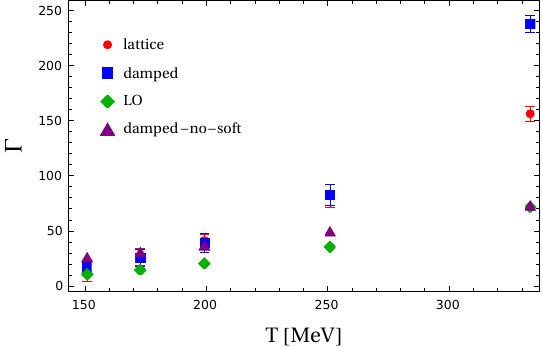}
\caption{The temperature dependence of the binding energy and thermal width (in MeV) of the bottomonium ground state.
The LO includes the one-loop static quark self-energies.  
\label{plot-peter}}
\end{centering}
\end{figure}

Next we compare with the results of \cite{Larsen:2019zqv}. We solve the Schr\"odinger equation using our result for the real part of the potential 
in the damped approximation.
We find the binding energy of the ground state at all temperatures available and subtract the Coulomb binding energy.\footnote{This subtraction makes the output independent of $S$.} In order to match the definition of Ref. \cite{Larsen:2019zqv},
the widths for each temperature are obtained by calculating the expectation value $-\langle {\rm Im}[V]\rangle$. 
The constants 
$(q_0,i_0,i_2)$ are determined by fitting to the lattice data at all available temperatures
and their values are 
$(q_0,i_0,i_2)=(0.044 \pm 0.002,-0.026 \pm 0.009 ,0.052 \pm 0.002)$. The errors are obtained from the fits to the upper and lower values. The results are shown in Fig. \ref{plot-peter}. If we ignore the soft contributions, we slightly improve (worsen) the description of the decay width (binding energy). When we include them, 
the temperature dependence of both the binding energy and the decay width gives a reasonable description of lattice data, and a considerable improvement with respect to the LO results. 

We can use our results for the binding energies and widths to estimate the dissociation temperature. 
The dissociation temperature of a bound state can be defined as the temperature for which its thermal decay width equals the energy difference to the closest state, at which point we are not able to distinguish  a given state from its neighbor in the spectral function. For simplicity, we will use instead the energy difference to the threshold, which is expected to provide a slightly higher dissociation temperature.
The decay width is defined to be 
the expectation value of $-2$Im$[V]$. The leading order result is 
$T_{\rm diss} = 193.2$~MeV and 
the damped approximation gives $T_{\rm diss} = 210 $ MeV if we ignore the soft contributions, $T_{\rm diss} = (202\pm 10)$ MeV using the fits to Ref. \cite{Larsen:2019zqv}, and $T_{\rm diss} = (142.7\pm 1.1)$ MeV using the fits to Ref.~\cite{Bazavov:2023dci} (see Fig.~\ref{plot-dis}). 

Our approach is able to reasonably describe the two sets of lattice data \cite{Bazavov:2023dci,Larsen:2019zqv}. However, the parameter sets we need in order to do so are not entirely consistent. We believe this might be due to a problem in the extraction of the decay width in either lattice paper. Since in our approach all scales are explicit, we expect the same size for  the numerical coefficients obtained from the two fits. This is true for all coefficients in the two sets except for $i_2$ in the set of Ref. \cite{Bazavov:2023dci}. The outcomes for $q_0$ differ by only a factor $\approx 0.61$, for $i_0$ by a factor $\approx 0.73$, but for $i_2$ they differ by a factor $\approx 3.7$.
The dissociation temperature we obtain from the set of Ref. \cite{Bazavov:2023dci} is also very low, incompatible with earlier lattice studies that indicate it is much higher than the crossover temperature; see for instance \cite{Aarts:2014cda}.

{\bf 4. Summary and conclusions.}
We have calculated the momentum space potential including corrections beyond the leading order HTL result, when the typical momentum transfer $p$ satisfies $m_D\ll p\ll T$. This is the relevant region  to obtain the dissociation temperature for heavy quarkonium. We have extended our calculation to softer $p$ by keeping suitable $p^2+m_D^2$ terms unexpanded, which we have called the damped approximation. 
We also include the power corrections calculated in QCD. The coordinate space potential is then determined for $1/T\ll r \ll 1/m_D$ up to a polynomial in $r^2$
, which encodes the contribution for $p\lesssim m_D$ at any order in $g$. We expect our expressions (\ref{extra-1}) and (\ref{extra-2}), with the addition of (\ref{soft}), to be reasonable approximations at larger $r$ as well, since explicit damping factors are kept. 
Hence, we expect they will provide useful inputs for the Bayesian methods required in the effort to determine the real-time static potential from lattice QCD. As an example, we have shown that our results describe reasonably well two different sets of lattice data, 
whereas the LO 
result fails to do so. We have also been able to identify an inconsistency between these two sets of data. This is not surprising because of the notorious difficulty of extracting the imaginary part of the potential in lattice calculations. Our results could be used as a check when comparing the validity of different methods.

\begin{acknowledgments}

We thank Peter Petreczky and Rasmus Larsen for providing the data of Refs. \cite{Larsen:2019zqv} and \cite{Bazavov:2023dci}, respectively. MEC acknowledges support by the Natural Sciences and Engineering Research Council of Canada under grant SAPIN-2023-00023 and thanks ICCUB and ICE for hospitality. CM  was supported by Ministerio de Ciencia, Investigaci\'on y Universidades (Spain) MCIN/AEI/10.13039/501100011033/ FEDER, UE, under the project  PID2022-139427NB-I00, by Generalitat de Catalunya by the project 2021-SGR-171 (Catalonia), and also partly supported by the Spanish program Unidad de Excelencia
 Maria de Maeztu CEX2020-001058-M, financed by MCIN/AEI/10.13039/501100011033.
JS acknowledges financial support from Grant No. 2017-SGR-929 and 2021-SGR-249 from the Generalitat de Catalunya and from projects No. PID2022-136224NB-C21, PID2022-139427NB-I00 and No. CEX2019-000918-M from Ministerio de Ciencia, Innovaci\'on y Universidades.

\end{acknowledgments}

\appendix

\section{Supplemental Material}

We discuss some of the technical steps needed for the computation of the high temperature QCD static potential beyond leading order. See \cite{Laine:2006ns} 
for details of the leading order calculation, \cite{Ghiglieri:2020dpq} 
for a recent review of perturbative thermal field theory, and \cite{Rothkopf:2019ipj}
 for a wider review that includes non-perturbative techniques.

For the retarded, advanced and time-ordered longitudinal gluon propagator we use the notation
\bea
&& G
^{\rm ret}(p_0,\vec p) =-\frac{1}{p^2+\Pi^{\rm ret}
(p_0,\vec p)} 
\ , 
\nn \\
&& G
^{\rm adv}(p_0,\vec p) = -\frac{1}{p^2+\Pi^{\rm adv}
(p_0,\vec p)} \ , \nn \\
&& G
(0,\vec p) = \frac{1}{2}\lim_{p_0\to 0}\big[(G^{\rm ret}
(p_0,\vec p)+G^{\rm adv}
(p_0,\vec p) \nonumber \\
&&+(1+2 n_b(p_0))(G^{\rm ret}
(p_0,\vec p)-G^{\rm adv}
(p_0,\vec p))\big] ,
\label{timeordered}
\eea
where $n(p_0)$ is a Bose-Einstein distribution and $\Pi^{\rm adv}
(p_0,\vec p)={\Pi^{\rm ret} 
(p_0,\vec p)}^\ast$.
In Coulomb gauge 
$\Pi^{\rm ret}
(p_0,\vec p)$ = $\Pi_{\rm HTL}^{\rm ret}(p_0,p)$ + $\Pi^{\rm ret}_{\rm BLO}(p_0,\vec p)$ where the first term is the HTL (leading order) result and the beyond leading order contribution is given in Eq. (4).

The potential is defined as
\bea
&& V(\vec r) = \lim_{t\to\infty} v(t,\vec r) \ , \label{bigV}\\
&& v(t,\vec r) = \frac{i}{t} \ln[C(t,\vec r)] \ ,\label{littlev}\\
&& C(t,\vec r) =  \frac{1}{N_c} \langle W_1 W_2 W_3 W_4 \rangle \ ,\label{bigC}
\eea
where the notation $W_i$ refers to the Wilson line corresponding to each of the four branches of the loop. 
For example the line along the top of the loop is
\bea
W_1 = {\rm Exp}\left(i g {\cal P}\int_{(0,\vec r)}^{(t,\vec r)} dx^\mu A_\mu(x)\right) \ ,
\label{W-def}
\eea
and the symbol ${\cal P}$ indicates path ordering. 
In the limit $t\to\infty$ the lines along the right and the left sides can be set to one. 
We calculate the potential perturbatively by expanding in $g$ to obtain
$
C_0(t,\vec r) + g^2 C_1(t,\vec r) + g^4 C_2(t,\vec r) + \dots$ =  exp$(-it\left(g^2 v_1(t,\vec r) + g^4 v_2(t,\vec r) + \dots\right))$. 
Matching orders in $g$ gives
\bea
 V_1(\vec r) &&= \lim_{t\to \infty} v_1(t,\vec r) = \lim_{t\to \infty} \left(\frac{i C_1(t,\vec r)}{t} \right) \ , \label{Vlo} \\
 V_2(\vec r) &&= \lim_{t\to \infty} v_2(t,\vec r)  \nonumber \\
&& = \lim_{t\to \infty} \left(\frac{i C_2(t,\vec r)}{t} - \frac{i C^2_1(t,\vec r)}{2t}\right)\,. \label{Vnlo}
\eea
To obtain the lowest order potential $V_1(\vec r)$ we calculate $C_1(t,\vec r)$ which has two contributions, one from the contraction of a gauge potential in the expanded link operator $W_1$ with a gauge potential from the expanded link operator at the bottom of the loop, and one from a contraction of two gauge potentials from the same line. These two contributions are called respectively $C_{1a}(t,\vec r)$ and  $C_{1b}(t,\vec r)$ and are
\begin{widetext}
\bea
&& C_{1a}(t,\vec r) = -i g^2 C_F  \int_0^t dx_0 \int_0^t dy_0 \int \frac{d^4p}{(2\pi)^4} 
e^{-i(p_0(x_0-y_0)-\vec p\cdot\vec r)}G_{00}(p_0,\vec p)\ , \\
&& C_{1b}(t,\vec r) = 2 g^2 C_F \int \frac{dl_0}{2\pi} 
\int_0^t dx_0 \int_0^{t} dy_0 e^{il_0(x_0-y_0)} \int \frac{d^4p}{(2\pi)^4}  \frac{1}{l_0+p_0 -i\eta} G_{00}(p_0,\vec p) \,.
\eea
\end{widetext}
To take the limit that the time goes to infinity we use
\bea
&& \lim_{t\to\infty} \int_0^t dx_0 \int_0^{t} dy_0 \, e^{il_0(x_0-y_0)} \nonumber \\
&& = t\,2\pi\delta(l_0) + \left(\frac{1}{(l_0+i\eta)^2} +  \frac{1}{(l_0-i\eta)^2}\right)\,. \label{X4}
\eea
To obtain this result we write 
\bea
\int_0^t dx_0 \int_0^{t} dy_0 \, e^{il_0(x_0-y_0)} = \frac{4}{l_0^2}\sin^2\left(\frac{tl_0}{2}\right)\,.
\label{D1}
\eea
It is straightforward to obtain the leading order term in (\ref{X4}). 
The time independent constant is found by multipling by $l_0$ regulating each term consistently. 
We note that the $t$-independent term in (\ref{X4}) is not needed to get the leading order potential but is important beyond leading order. 

To find the potential at leading order in the expansion of the Wilson lines in $g$, we substitute $C_1(t,\vec r)=C_{1a}(t,\vec r)+C_{1b}(t,\vec r)$ into (\ref{Vlo}) and use (\ref{X4}).  This gives the result for $V_{1 {\rm LO}}$  in Eq. (2)
.
The expression for $V_2$ is obtained from (\ref{Vnlo}) and can be written in the form
\bea
V_2 &&= \lim_{t\to\infty} \bigg[
\left\{\frac{i C_{2{\rm b}}}{t} - \frac{i}{2t} [C_{1a}(t)]^2\right\}
+\frac{i C_{2{\rm c}}}{t} \nonumber \\
&&+\frac{i C_{2{\rm d}}}{t} 
+\left\{\frac{i C_{2{\rm e}}}{t} - \frac{i}{t} C_{1a}(t)C_{1b}(t) \right\}\bigg]\,.
\label{curly-2}
\eea
The role of the factors $C_{1a}(t)$ and $C_{1b}(t)$ that appear in Eq. (23) 
is to cancel the pinch singularities that appear in Figs. 1 
(b)(e). 
A straightforward calculation gives
\bea
V_2^{\rm (bc)} (r) & \equiv& V_{2}^{\rm (b)}(r)+V_{2}^{\rm (c)} (r) \\
&= &
 \lim_{t\to\infty} \left[
\left\{\frac{i C_{2{\rm b}}}{t} - \frac{i}{2t} [C_{1a}(t)]^2\right\}
+\frac{i C_{2{\rm c}}}{t} 
\right] \nn 
\\
&=& -i\frac{g^4 N_c C_F}{2} 
\int\frac{d^3p}{(2\pi)^3} e^{i \vec p \cdot \vec r}\nonumber \\
& \times& \int\frac{d^4k}{(2\pi)^4} 
\frac{ G(k_0,\vec k+\vec p) G(k_0,k) }{(k_0+i\eta)^2}\nonumber
\ , \label{ladd-cross-all-2}\\
V_2^{\rm (de)} (r) & \equiv& V_{2}^{\rm (d)}(r)+V_{2}^{\rm (e)} (r) \\
&=& \lim_{t\to\infty} \left[
\frac{i C_{2{\rm d}}}{t} 
+\left\{\frac{i C_{2{\rm e}}}{t} - \frac{i}{t} C_{1a}(t)C_{1b}(t) \right\}\right] \nn \\
&
= & ig^4 N_c C_F \int \frac{d^3l}{(2\pi)^3}e^{i\vec l \cdot \vec r}G(0,l) \nonumber \\
&& ~~~~~~~~~~~~ ~~~ \times \int \frac{d^4k}{(2\pi)^4}\frac{G(k_0,k)}{(k_0+i\eta)^2}\,.\nonumber
\label{vert-wave-all-2}
\eea

The calculation described above has been cross-checked against an independent calculation of the diagrams of Fig. 1 
carried out by standard diagrammatic techniques.

\begin{figure}[h]
\label{de}
\begin{center}
\includegraphics[scale=0.8]{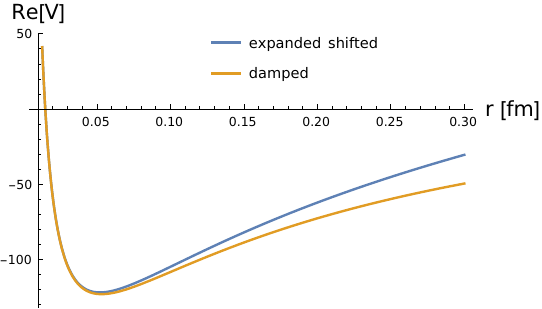}
\\ 
\end{center}
\begin{center}
\includegraphics[scale=0.8]{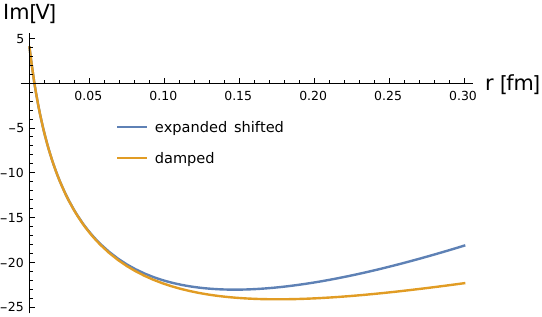}
\end{center}
\caption{We compare $V=V_2$ (in MeV) at $T=182$ MeV in the damped approximation (10)-(11) (orange), and in the (minimally subtracted with $\mu=m_D$) expanded result (\ref{expanded}) shifted by a constant so that they match at $r=0.01$ fm (blue).}
\end{figure}

We give here the result of the Fourier transform of the fully expanded ($m_D\ll p \ll T$) momentum space potential (9) in DR for $d\to 3$,

\bea
&&V_{2, {\rm exp}}(r)=-\frac{g^4C_F N_c T}{16\pi m_D}\left\{\left[
\left(1-\frac{3 \pi ^2}{16}\right)\frac{1}{4 \pi r}\right.\right. \nonumber \\
&&\left.\left.-  \frac{ m_D}{ \pi} L(r)
- \left(\frac{5\pi ^2}{24}-\frac{4
   }{3}\right)\frac{r m_D^2}{8\pi}\right]\right. 
	\\
&&\left. -i\pi T m_D \left[ 
\frac{7r}{3 \pi^2} -\left(1-\frac{3 \pi ^2}{16}\right) \frac{m_D r^2}{24\pi^2}\left(1-L(r)
\right)\right.\right.\nonumber\\ 
&& \left.\left.
-\frac{1}{\pi^3 T}\left(1-\frac{N_f}{2N_c}\right)
L(r)
\right]\right\}\nonumber\,.
\label{expanded}
\eea
$L(r)=-2/(d-3)+\gamma+\log\left(\pi (r\mu)^2\right)$. The $1/(d-3)$ poles can be absorbed in the parameters of $V_{2,{\rm soft}}$ in (12). The damped approximation regulates these poles. Namely, it reshuffles part of the soft contribution into the semi-hard one. In Fig. \ref{de} we compare the damped approximation with the fully expanded result for $V_2$. The fact that a shift makes them agree at short distances is consistent with (12).

\vspace{0.2cm}

Next we prove that the uncalculated soft contributions have the form given in Eq. (12) at the order to which we work.
Consider
\bea
V(r) = \int \frac{d^3p}{(2\pi)^3}e^{i\vec p\cdot\vec r}\tilde V(p)\,, \nn
\eea
with $r\sim g^{-a}T^{-1}$, $1/3<a<2/3$. 
 For soft momenta we can take advantage of the fact that $pr\sim m_D r \sim g^{1-a}<1$ to expand the exponential which gives
\bea
V_{\rm 
soft}(r) &=& \int \frac{d^3p}{(2\pi)^3}\left(1+ i\vec p\cdot\vec r -\frac{1}{2} (\vec p\cdot\vec r)^2 +\cdots\right)\tilde V(p) \nonumber\\
&\equiv& V_{\rm soft}^0+V_{\rm soft}^2(r) +\cdots
\,.\label{V-soft}
\eea
The second term is zero by symmetry in an isotropic system. Note that an UV regulator may have to be introduced and the result would then be scale dependent (we use dimensional regularization so that all integrals are defined). 
If $p$ is soft then the biggest contribution to the imaginary part of $\tilde V(p)$ from the diagrams we have calculated is $\sim g^4 T^2/p^4$ where the $T^2$ is from two Bose enhancement factors and the $p^{-4}$ follows from dimensional analysis. The biggest contribution to the real part of $\tilde V(p)$ from the diagrams we have calculated is $\sim g^4 T/p^3$ because there can be only one Bose enhancement factor. 
From these results we have that 
\bea
 {\rm Im} V_{\rm soft}^0 &=&p^3  {\rm Im}[\tilde V]_{\rm soft}(p) \sim  \frac{g^4 T^2 }{p} \sim g^3 T \nn \\
{\rm Im} V_{\rm soft}^2 &=& r^2 p^5  {\rm Im}[\tilde V]_{\rm soft}(p) \sim r^2 g^4 T^2 p \nn\\
& \sim & r^2 g^5 T^3 \sim g^{5-2a} T\nn \\ 
 {\rm Re} V_{\rm soft}^0 &=& p^3  {\rm Re}[\tilde V]_{\rm soft}(p) \sim  g^4 T  \,,
\label{keep}
\eea
where we have used $r\sim g^{-a}T^{-a}$.
To determine which of these terms should be included,  we compare them with the terms we have dropped in our analytic calculation of the semi-hard contributions. 
As explained 
above Eq. (9),
we have dropped
\bea
&& \text{Im}V_2: \,
\left(\frac{g^4 T^2 m_D^2}{p^6},\frac{g^4}{p^2}\right) 
= \left(g^{6-3a}T,g^{4+a}T\right)
\label{dropi} \\
&& \text{Re}V_2: \,
\frac{g^4 T m_D^2}{p^5} = g^{6-2a}T\,
\label{dropr}
\eea
(for the imaginary part the first factor is bigger for $1/2<a<2/3$ and the second for $1/3<a<1/2$). 
Comparing Eqs. (\ref{keep}), (\ref{dropi}), (\ref{dropr}) we see that all terms in (\ref{keep}) should be kept,
and higher order terms starting at 
${\rm Im} V_{\rm soft}^4$ and  ${\rm Re} V_{\rm soft}^2$  should not be included. This leads us to the terms displayed in (12).
 \\

\begin{figure}[h]
\begin{centering}
\includegraphics[scale=0.88]{
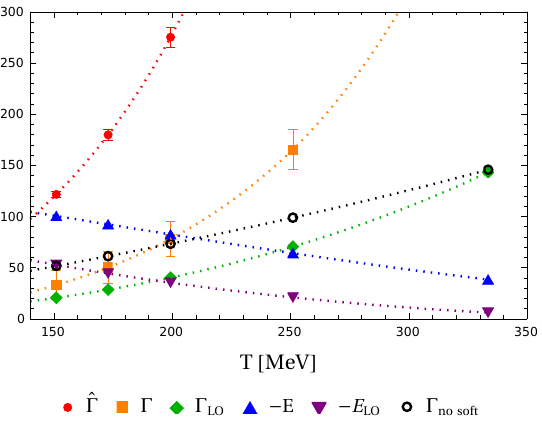}
\caption{
The binding energy ($-E$) and the thermal width ($\Gamma$) as a function of temperature. The dotted lines are interpolated curves that are added to guide the eye.  
$\Gamma_{\rm no\,\, soft}$, $\Gamma$ and $\hat\Gamma$ indicate the curves obtained with no soft contribution, and with soft contribution using the fits to Ref.  [43] 
and Ref. [16] 
respectively. The LO results include the one-loop static quark self-energies.  
   \label{plot-dis}}
\end{centering}
\end{figure}

Finally, we illustrate in Fig. \ref{plot-dis} how we obtain the dissociation temperatures.

{}

\end{document}